\documentclass{egpubl}
 
\JournalPaper         
\usepackage[T1]{fontenc}
\usepackage{dfadobe}

\usepackage{xcolor}
\usepackage{color}
\usepackage{enumerate}
\usepackage{textcomp}
\definecolor{orange}{rgb}{1,0.5,0}

\usepackage[T1]{fontenc}
\usepackage{tabu}                      
\usepackage{booktabs}       
\usepackage{blindtext}

\biberVersion
\BibtexOrBiblatex
\usepackage[backend=biber,bibstyle=EG,citestyle=alphabetic,backref=true]{biblatex} 
\electronicVersion
\PrintedOrElectronic

\ifpdf \usepackage[pdftex]{graphicx} \pdfcompresslevel=9
\else \usepackage[dvips]{graphicx} \fi

\usepackage{egweblnk} 

\title[EG \LaTeX\ Author Guidelines]%
      {PDViz: a Visual Analytics Approach for State Policy Data \vspace{-1.4em}}

\author[Dongyun Han, 
Abdullah-Al-Raihan Nayeem, 
Jason Windett
\& Isaac Cho]
{\parbox{\textwidth}{\centering
Dongyun Han$^{1}$, 
Abdullah-Al-Raihan Nayeem$^{2}$,
Jason Windett$^{2}$,
and Isaac Cho $^{1, 2}$
}
        \\
{\parbox{\textwidth}{
\centering $^{1}$ Utah State University, Utah, USA  \\ 
            dongyun.han@usu.edu, isaac.cho@usu.edu\\
         $^{2}$University of North Carolina at Charlotte, North Carolina, USA\\
            anayeem@uncc.edu, jwindett@uncc.edu
         \vspace{-3em}
       }}
}

\begin{document}

\teaser{
 \includegraphics[width=\linewidth]{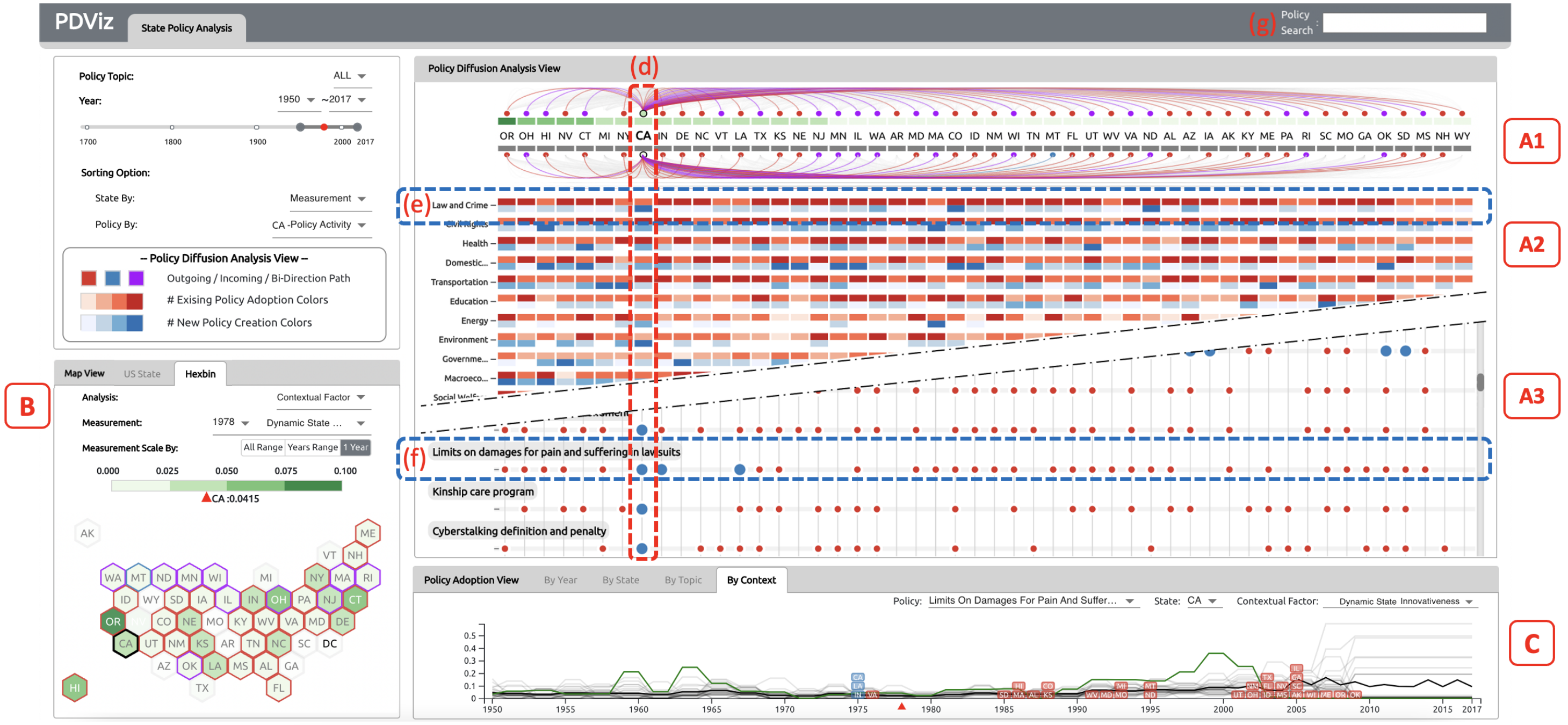}
 \centering
  \caption{PDViz overview. The interface has three views: Policy Diffusion Analysis (A1, A2 and A3), Geographic (B), and Policy Adoption Views (C). The policy diffusion analysis view has the policy diffusion chart (A1) and policy matrix chart (A2 \& A3). The drop-down menus allow the user to select multiple options to analyze the policy diffusion data.
  The system help the user analyze the policy data.
  }
\label{fig:teaser}
}

\maketitle
\begin{abstract}
   Sub-national governments across the United States implement a variety of policies to address large societal problems and needs. Many policies are picked up or adopted in other states. This process is called policy diffusion and allows researchers to analyze and compare social, political, and contextual characteristics that lead to adopting certain policies, as well as the efficacy of these policies once adopted. 
In this paper, we introduce PDViz, a visual analytics approach for social scientists to dynamically analyze the policy diffusion history and underlying patterns.
It is designed for analyzing and answering a list of research questions and tasks posed by social scientists in prior work. To evaluate our system, we present two usage scenarios and conduct interviews with domain experts in political science. The interviews highlight that PDViz provides the result of policy diffusion patterns that align with their domain knowledge as well as the potential to be a learning tool for students and researchers to understand the concept of policy diffusion. 
\keywords{Visual Analytics, Political Science, Policy Diffusion}

\ccsdesc[500]{Human-centered computing~Visual analytics}
\ccsdesc[500]{Information systems~Information systems applications}

\printccsdesc   
\end{abstract}  
\section{Introduction}
In the United States' (US) federated system of government, both state and federal policies influence the welfare, education, economic outlook, and safety of everyone within the nation's borders.
They govern where businesses can locate, how elections are administered, whether to raise the minimum wage, and a host of other issues that impact our daily lives. Because states implement a variety of policies, there is significant variation in how governments deliver services and address community problems. This diversity offers tremendous potential for governments to innovate and learn from the creative approaches of others. It also allows researchers to explore the efficacy of a wide range of policies.

Through continuous research, great progress has been made in understanding policy innovation and diffusion. Both areas of scholarship are concerned with the willingness of US states to adopt or create innovative policies sooner or later than other states to respond to a social need or an emerging issue within or between states \cite{boehmke2012state}. Desmarais et al.~\cite{desmarais2015persistent} advanced our understanding of diffusion and innovation by using a network inference algorithm that can characterize latent underlying policy diffusion patterns across states and over time. These studies identifying states' tendencies and pathways \cite{walker_diffusion_1969, gray_1973} in the perspective of policy diffusion are crucial as states repeatedly face similar political situations to estimate future adoptions (see \cite{mooney_2021} for a larger discussion of this literature). However, many still take a static approach to analyzing them in spatio-temporal and policy contexts~\cite{traunmuller2020visualizing}.


In this paper, we introduce PDViz, a visual analytics approach that allows the user to dynamically analyze the diffusion of policy in social and political contexts. 
To determine the specifications, we compile a list of research questions and tasks posed by political scientists in previous works~\cite{boehmke2020spid, ahn2020policyflow, boehmke2012state}. We utilize the latest State Policy Innovation and Diffusion Database (SPID)~\cite{boehmke2020spid} and Correlates of State Policy Project (CSPP)~\cite{grossmann2021correlates} datasets. They are comprehensive datasets collected from a variety of sources including academics~\cite{boushey2016targeted, makse2011role, caughey2016dynamics} and non-profit governmental organizations~\cite{ULC, NCSL}. SPID is inherently designed for estimating policy innovation and tracing diffusion ties while CSPP is designed for executing social research. 
PDViz has coordinated multiple views~\cite{andrienko2007coordinated} to show different features of the database and provide a better understanding of its context and patterns to a wide range of users, including political scientists and policymakers. We present two usage scenarios and an evaluation of interviews with domain experts. The evaluation is conducted on the basis of PDViz's ability to support in performing analytical tasks. The results indicate that PDViz is an effective visual analytics tool that allows the users to explore policy adoption records and analyze the underlying diffusion patterns. Furthermore, it shows the potential to be a tool for education in secondary and postsecondary settings to help students understand historical and contemporary policy debates, outcomes, and the social and political context that led to adoption (or not). The main contributions of this work are described as follows.


\begin{enumerate}
\item We introduce PDViz that allows the user to analyze the diffusion patterns of policies in relation to social and political factors.
\item We provide a pipeline for our data-driven approach to analyze and visualize the diffusion patterns of policy with geographical, temporal, and contextual factors.    
\item  We present two usage scenarios and interviews with domain experts to evaluate the PDViz system.
\end{enumerate}

\section{Related Work}
\subsection{Visual Analytics Systems for Political Science}

Political scientists analyze social relationships and influences and share their insights with policymakers to create and revise policies for a sustainable future. 
Data visualization approaches have been applied for public policy analysis~\cite{kohlhammer2012toward, nazemi2014semavis, Stimson1991}, public health~\cite{tariq2021planning, livnat2012epinome}, public safety~\cite{zanabria2019crimanalyzer}, economy~\cite{wang2018visual}, and traffic~\cite{chen2015interactive}. 
For example, Kohlhammer et al.~\cite{kohlhammer2012toward} applied visualization approaches to support policy-making processes. They introduced a simplified policy life cycle for visualization researchers and present ePolicy (Engineering Policy Making Life Cycle)~\cite{epolicy} and Fupol (Future Policy Modeling)~\cite{fupol} as its usage cases.

Understanding the potential and dynamic interdependencies between society members is one of the major topics in the political science field. Understanding policy diffusion and political public opinion are included in it. Much research has been done on policy diffusion, as described in Section~\ref{section_policy_analysis_background}, but only a few visual analytics approaches have been explored. 
Ahn and Lin presented PolicyFlow~\cite{ahn2020policyflow}, a visualization interface for the user to analyze the policy diffusion patterns in geospatial and temporal contexts as well as political contexts such as policy topics and similarities. 
They provided a novel matrix-based diffusion representation showing an overview of how source and target states are mapped to each other, and whether each adoption case falls into expected patterns.


PDViz has many key differences compared to PolicyFlow. First, PDViz allows the user to compare the policy diffusion patterns at two different levels, whereas PolicyFlow only allows the user to analyze a single policy pattern. On top of that, PDViz supports the user's exploratory data analysis of policy adoption records and policy diffusion in multiple contexts. Despite the fact that PolicyFlow includes additional views such as a pie chart and map view, their roles are limited to filtering. PolicyFlow highly depends on users' domain knowledge to use it. Third, PDViz actively adopts a larger choice of covariates and methods used in political science. The used covariates represent more appropriate predictors of policy spread, rather than a handful of highly correlated population characteristics. We utilize these variables to fit a model like the Cox Proportional-Hazards model to show the estimated effects of each covariate in a fully specified model, while PolicyFlow shows simply bi-variate correlations. Finally, PDViz provides an evaluation of the political scientists who have been working on research in SPID.

\subsection{Visual Analytics Systems for Information Diffusion}
Information diffusion is the process by which information spreads from one place to another. This includes the spread of rumors, memes, and ideas that could be spread differently by space, time, and subject through social networks~\cite{wu2014opinionflow, xu2018vaut, wang2016ideas, sahnan2021diva, Liu17b}.
Visual analytics approaches have been studied for many years to find insights~\cite{cho2017crystalball, zhao2014fluxflow} and understand causality patterns~\cite{wang2018visual}. For example, Karduni et al. introduced Verify ~\cite{karduni2018can, karduni2019vulnerable}, a visual analytics system that allows the user to detect misinformation and analyze how misinformation is distributed within social networks. Li et al.~\cite{li2018weseer} introduced WeSeer, which predicts the future popularity of social media posts and enables the user to understand the factors that contribute to popularity by providing the degree of propagation rate through the social network. The systems successfully showed their potential abilities to help the user understand patterns and complete given tasks. However, they limitedly support the user to compare the patterns in different levels (i.e., general vs. topic-specific patterns) between entities in a network. PDViz adopts an arc diagram layout to present two different underlying patterns up and down for the user to compare them.

\vspace{-10pt}
\subsection{Studying Policy Adoption History and Network Dataset}\label{section_policy_analysis_background}

Central to the analysis of public policy diffusion are the patterns, pathways, and mechanisms underlying policy spread~\cite{jensen2003policy, graham2013diffusion, boushey2010policy}. Desmarais et al.~\cite{desmarais2015persistent} adopted the network inference algorithm to political science, NetInf~\cite{gomezrodriguez_inferring_2010}, that is a machine learning algorithm that tracks diffusion patterns through networks and infers the underlying diffusion network.  Desmarais et al. showed that NetInf can characterize the latent underlying influence of policy diffusion between states. They also created and distributed an R package named NetworkInference~\cite{netinf}. 
 The package contains the SPID dataset with 724 different policies coded by topic areas~\cite{boehmke2020spid} and NetInf.

Node centrality is also a good metric to measure relative importance of nodes in network data.
It includes PageRank and degree and closeness centrality. They are utilized to find the most influential people on social networks and scholars in the previous work~\cite{bollen2006journal, zhao2014fluxflow, li2013visual, page1999pagerank}. The centralities in the policy diffusion network allow the user to analyze which states play an important role in the history of policy adoption (see Section~\ref{Policy_Analysis_Measurements} for the details.)

The innovativeness score measures a state's willingness to adopt policies sooner or later than others. Walker~\cite{walker_diffusion_1969} first studied how policies travel from the state that first makes a policy to the rest of the US states and introduced static innovation scores. However, it has been pointed out that the score has right-censoring issues. Boehmke and Skinner~\cite{boehmke2012state} introduced a dynamic measure of innovation to address it, called dynamic state innovativeness. A state having a higher score indicates that it is more likely to adopt policies faster and is more innovative than other states.  

The Cox's proportional-hazard model (Cox model)~\cite{cox1972regression} is a regression model commonly used to address death/survival in medical treatment. This technique has been applied to model the probability of policy adoption or not across states over time~\cite{jones2005beyond}. We describe more details in Section~\ref{data_processing_and_analysis}.

\section{PDViz: Visual Analytics System for Policy Data}

The team of visualization researchers and a political scientist (the third author, one of the co-PIs of SPID) met biweekly for 1 year and had multiple iterations including task analysis, data processing and analysis, and visual interface design~\cite{meyer2019criteria, holtzblatt1997contextual}.

\subsection{Task Analysis}
We first started by reviewing the literature to identify common questions and tasks that political scientists have in the study of policy diffusion. We identified four major tasks T1-T4 as follows: 

\begin{enumerate}[T1]
\item \textbf{Summarize policy adoption records and diffusion patterns~\cite{boehmke2020spid}.} 
As policy diffuses over time, there will be lots of connections between states, making it a complex network. Thus, providing a clear and simplified overview is critical to answering questions such as ``Which states serve as sources of the policy diffusion for other states?'' and ``What are the statistics for the policy adoptions by year, states, and topic?''
\item \textbf{Explore policy diffusion in geo-spatial and temporal context~\cite{boehmke2012state, caughey2016dynamics, walker_diffusion_1969}.} 
Political scientists have interests in exploring and identifying policy spread in geo-spatial and temporal contexts. Specific questions include ``Do innovative states cluster into regions of innovation?'' and ``How are impacts of the states in the policy diffusion examined over time?''
\item \textbf{Explore and compare policy diffusion in topical context~\cite{boehmke2020spid, boushey2016targeted}.} 
They have been interested in identifying which state is the policy leader (or laggards) across different policy areas and how each state responds to new waves of policies being adopted. The example questions are ``How does policy leadership vary across substantive policy areas?'' and ``How are diffusion patterns different by policy topics?''
\item \textbf{Analyze policy diffusion in Social and Political Factors~\cite{boehmke_disentangling_2004, berry_state_1990}.} Analyzing policy diffusion while considering factors like percentage of \textit{foreign born} and \textit{crime rate} has been devoted by political scientists. Questions such as ``Are some states in the US more innovative than others?'' 
and ``How can correlated factors (e.g., demographic attributes of the states) for diffusion be analyzed?'' are included in this task.
\end{enumerate}

We design PDviz to support all tasks (T1, T2, T3, and T4) while previous tools support only some tasks~\cite{brehmer2013multi}. For example, PolicyFlow only partially supports T2 and T3.

\subsection{Data Processing and Analysis}\label{data_processing_and_analysis}
\subsubsection{Policy Diffusion Episode Data \& Patterns}
For our visual analytics tool, the latest SPID (version 1.2) is used. It has main policy diffusion data and separate meta-data in the csv format~\footnote{These two files are downloadable through the NetworkInference R package~\cite{NETWORKINFERNCE_R_CRAN} or Dataverse~\cite{SPID_DATAVERSE}}. The policy diffusion is a network data set, with each record describing which policy is spreading from one US state to another. It includes a total of 724 policies with twenty topics that spread across the 50 US states from 1691 to 2017 along with the 17,824 policy adoption records. Among the twenty topics, the two topics \textit{Foreign Trade} and \textit{Technology} are excluded because none of the policy diffusion events were classified in these areas during the collection of the SPID data. This coding schema is adopted from the Policy Agenda's Project Master Codebook \cite{cookbook}. These topics are mostly national-level projects that have issue areas that subnational units do not actively pursue legislation.  As a result, there remain 730 policies and 17,725 records.

We infer underlying policy diffusion patterns in the policy diffusion network. 
The patterns depict the common pathways of policy spread between states. 
Simultaneously, the patterns are used as predictors; for example, a pattern linked from state $i$ to state $j$ describes that state $j$ is more likely to adopt a policy in the near future if state $i$ recently adopted that same policy. To find the patterns, the NetInf algorithm in the NetworkInference R package~\cite{netinf} is utilized. 
It considers all the policy adoption records in SPID as a network, with the US states acting as nodes, and each policy adoption acting as a directed edge between two states. NetInf returns a list of edges inferred in each iteration until the Vuong test~\cite{vuong1989likelihood} for edge addition reaches the $p$-value cutoff. We choose 0.05 for the cutoff value and use the default values for all other NetInf input parameters except the input network. When the input network is SPID, it outputs an underlying diffusion network with 686 edges.

\subsubsection{Policy Analysis Measurements}\label{Policy_Analysis_Measurements}
The user can catch roles or impacts of the US States by observing their characteristics in the diffusion network. 
PDViz provides three methods to identify the characteristics, and each method has different measurement types as shown in Table~\ref{table_analysis_measurement_in_PDViz}. 
This subsection introduces \textit{Network Centrality} and \textit{Static State Innovativeness}, and the following subsection discusses the contextual factor.

PDViz provides five centrality measurements (\textit{Degrees, In-degrees, Out-degrees, Closeness, and PageRank}) over the underlying diffusion network. The \textit{Degree} centrality helps the user identify which states have actively influenced and been influenced across the policy diffusion network. Similarly, \textit{In-degree} shows how much a state has been influenced by others and \textit{Out-degree} shows how much a state has impacted others. \textit{PageRank} is useful for determining the importance of states. \textit{Closeness} allows the user to identify the states that play the most effective role in disseminating policies within the policy diffusion network. To measure them, we used NetworkX~\cite{hagberg2008exploring} with default input parameters.

The \textit{Static State Innovativeness} score shows a state's willingness to adopt policies sooner or later than others~\cite{boehmke2012state}. A state with a higher score is considered innovative and it is more likely to adopt new policies faster compared to the others. It can be measured over arbitrary time periods by adding total adoptions from year $Y\textsubscript{0}$ to $Y\textsubscript{1}$ and dividing by total annual adoption opportunities over the same time span. PDViz calculates the score for each state by following the user-defined topic and periods shown in Fig.~\ref{fig_option1}-a and -b.

\begin{table}[t]
    \includegraphics[width=1\columnwidth]{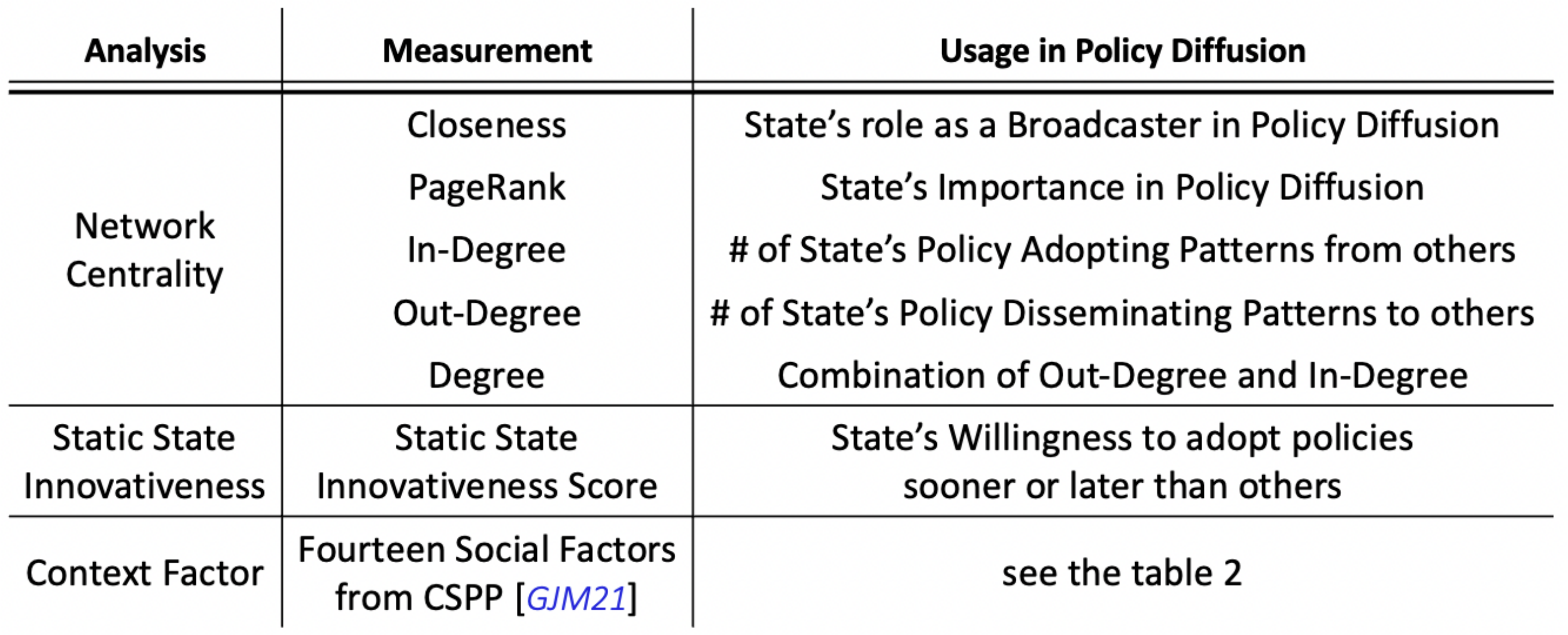}
    \centering
    \caption{Analysis methods and Measurements in PDViz}
    \vspace{-0.6cm}
\label{table_analysis_measurement_in_PDViz}
\end{table}

\subsubsection{Contextual Factors in Political Science}
The Correlates of State Policy Project (CSPP) dataset is used to analyze the policy diffusion from the perspective of social factors in PDViz.
It includes more than 2000 variables for the political, social, or economic factors from 1900 and 2019. We choose the 14 factors among them, as listed in Table~\ref{table_factors_in_CSPP}. These factors offer a holistic view of the context in which policies are developed both within and across states. Please note that the \textit{Dynamic State Innovativeness} score differs from the static score that we described earlier. The dynamic score varies by year despite the fact that both scores are measures of a state's propensity to adopt policies sooner or later than other states.

The 14 attributes from 1900 and 2017 are used because 2017 is the last year in SPID. However, not all attributes in the range have valid values, as some values are missing or not applicable. For example, the percentage values of \textit{Foreign Born} and \textit{African American} are collected by decade. We use the collected values from the previous decade for each US state until the following decade; for example, the years 2011-2017 have the same values as 2010. Another example is Nebraska, a unicameral state which has a non-partisan legislature. The numbers of senate and house democrats for Nebraska are not applicable. We filled 0 for this case.  Other missing values are filled in with the most recent non-missing value. 

\begin{table}[t]
    \includegraphics[width=\columnwidth]{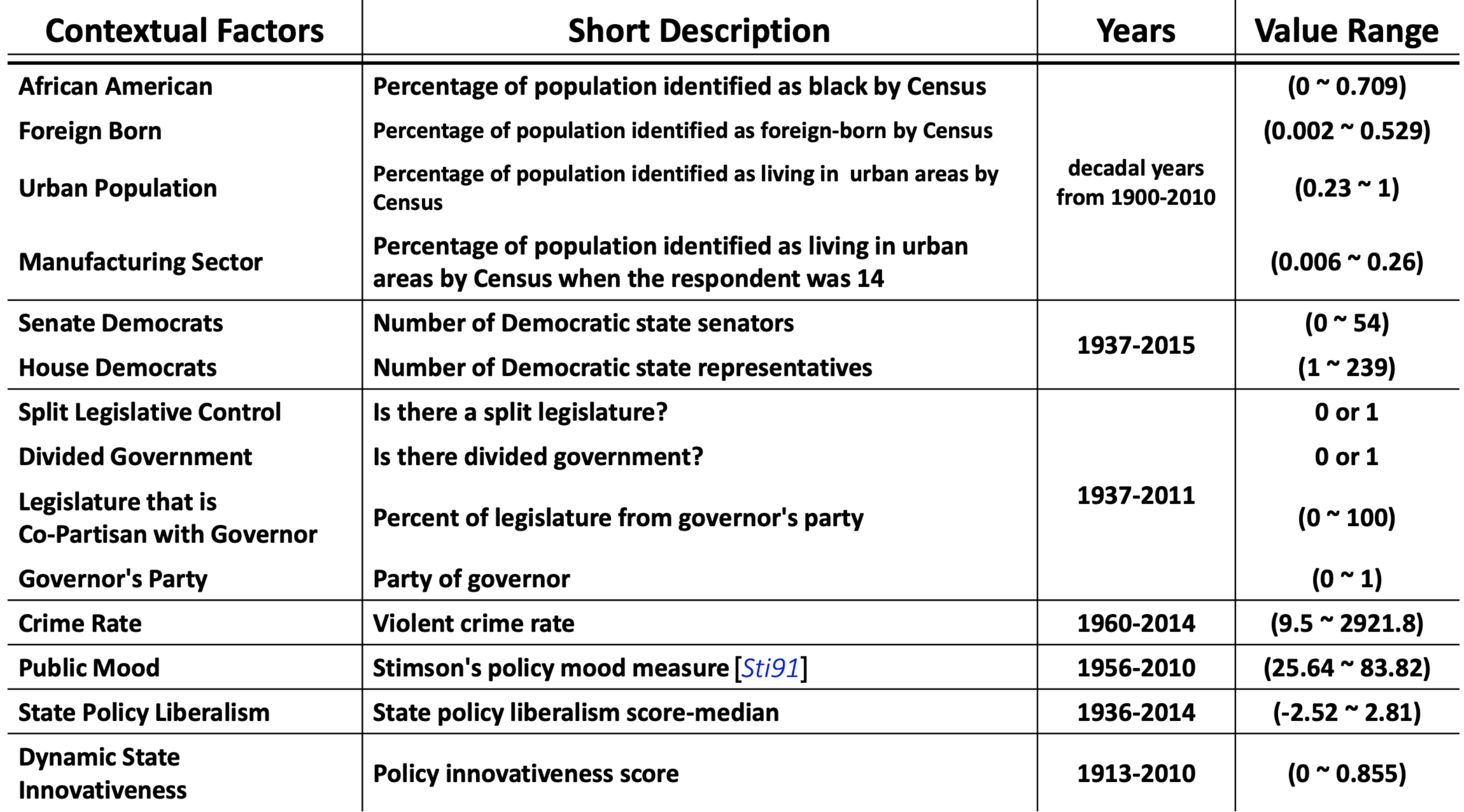}
    \centering
    \caption{ {\small Fourteen Social and Political Factors from CSPP~\cite{grossmann2021correlates}}}
    \vspace{-0.5cm}
\label{table_factors_in_CSPP}
\end{table}

\begin{figure*}[t]
  \centering
    \includegraphics[width=0.88\linewidth]{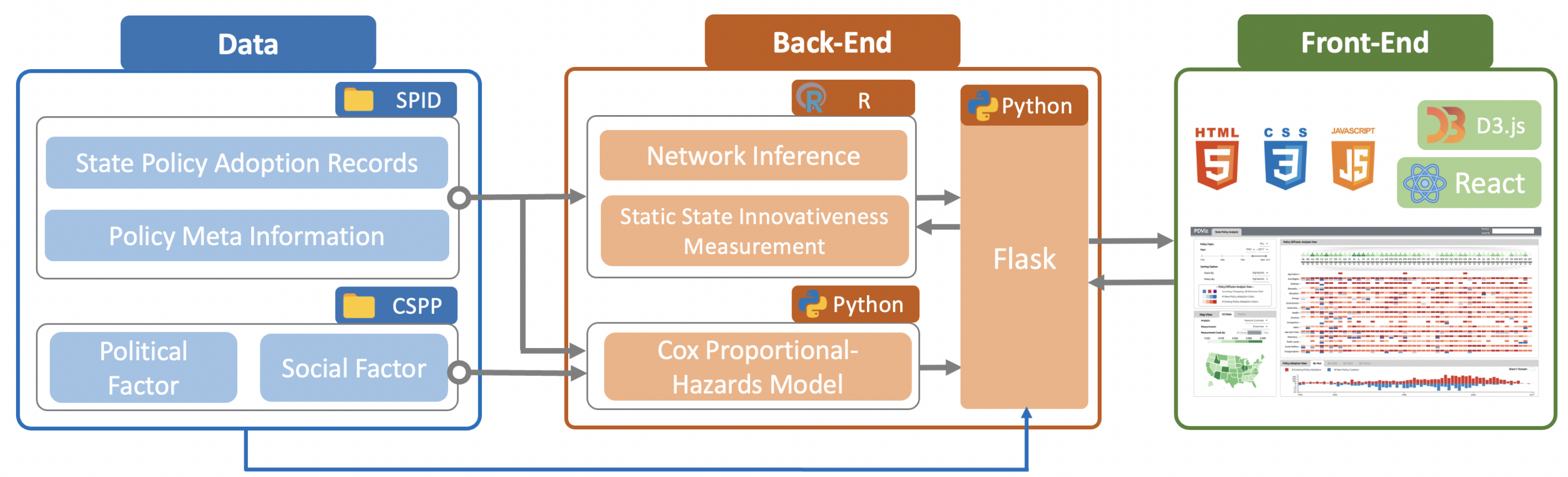}
     \caption {
    Data Flow Pipeline and Processes. PDViz is a data driven interface, and utilize the SPID and CSPP data implemented in domain of political science. 
    PDViz shows the processed results in the back-end by following the user's interactions in the front-end.
    } 
    \label{fig_pipeline}
    \vspace{-0.5cm}
\end{figure*}

\subsubsection{Cox Model \& Hazard Ratios}
Political scientists are interested in understanding the impact of the context factors on policy adoption. The Cox model is the widely used method to model the probability of policy adoption following the context factors over time~\cite{jones2005beyond}. For the Cox model, we utilize a Python library, Lifelines\cite{davidson2019lifelines}. We fit the Cox model using the input we organized from SPID and CSPP with a 95\% confidence interval for coefficient significance. The number of rows in the input data for each policy and state is equal to the years between a policy's first adoption year and the state's adoption year. 
Each row includes a year, the values of the social factors, and a true or false value indicating whether states adopted policies or not in the year. If a state has not adopted a policy until the policy's last adoption year, the last year is used as the state's adoption year. It outputs the hazard coefficients and ratios, $p$-values, etc. The expected probability of policy adoption (hazard) at time t is defined by: 
\vspace{-0.2cm}
\begin{equation}
    h(t) = h_{0}(t) exp (\sum_{i=1}^{n}\beta_{i}x_{i}), 
\vspace{-0.1cm}
\end{equation}
where $h_{0}(t)$ is the baseline probability, $n$ is the number of context factors, and $x$\textsubscript{$i$} and $\beta$\textsubscript{$i$} are a context factor and its hazard coefficient. The hazard ratio of a factor $x$\textsubscript{$i$} is defined as $exp$($\beta$\textsubscript{$i$}). The ratios show the relative impacts of the context factors to the adoption probability. When the ratio of a factor is bigger than 1, it indicates that the factor positively affects US states to adopt the policy.

$P$-values indicate the relationships between the factors and the possibility of policy adoption. The lower the $p$-value, the stronger the relationship between a factor and an adoption. For example, the policy \textit{Laws establishing Hate Crimes against Minorities (Fig.~\ref{fig_usage_scenario2}-e)} has about 49.9 and 16.9 hazard ratios for \textit{Dynamic State Innovativeness} and percent of \textit{Foreign Born} with .047 and .036 $p$-values respectively. When all other factors are held constant, it indicates that US states with higher values of the two factors have a higher chance of adopting the policy. The $p$-values show that the \textit{Foreign Born} factor is a stronger indicator than the innovativeness score.

\begin{figure}
  \centering
    \includegraphics[width=.9\columnwidth]{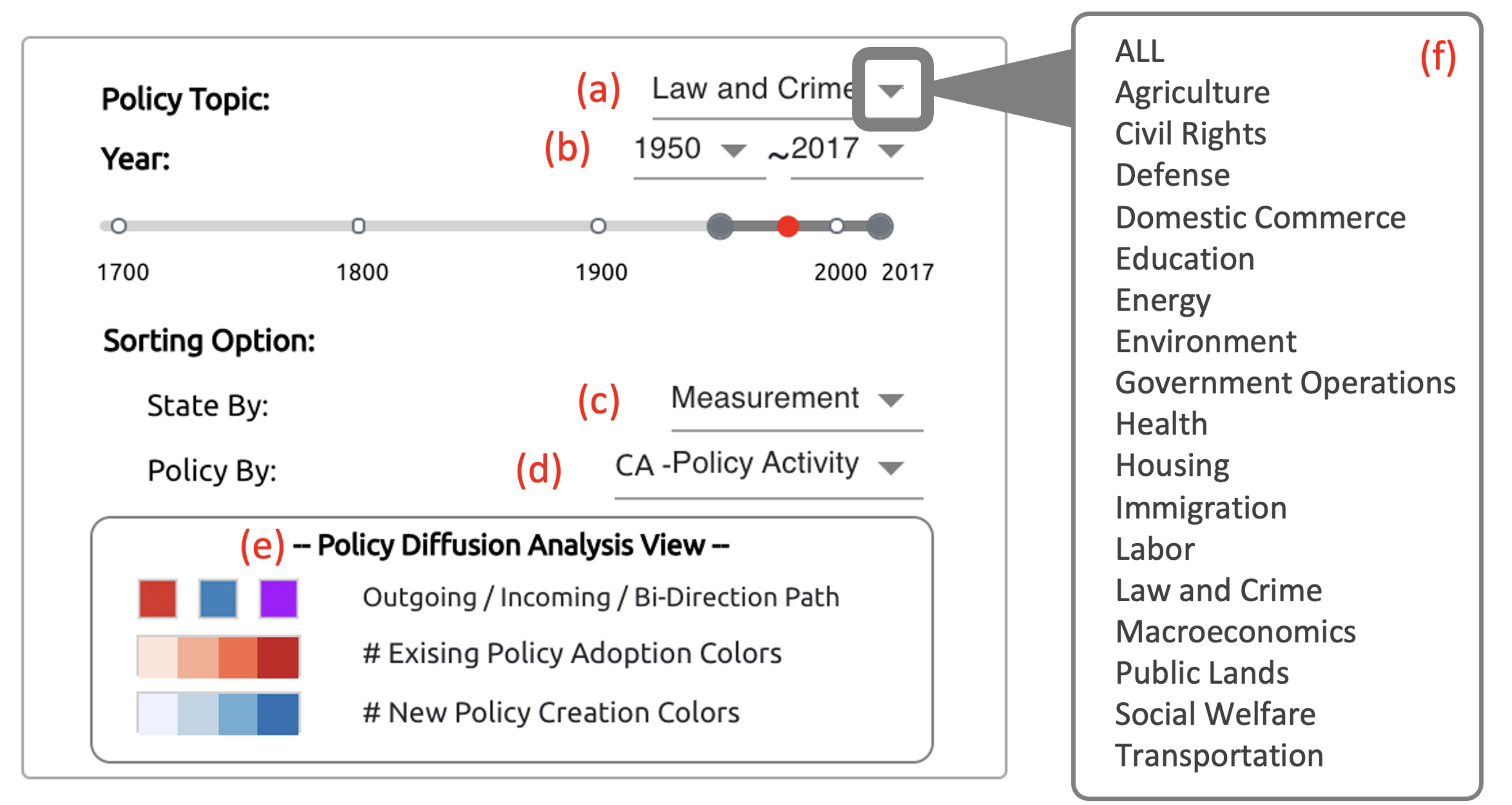}
     \caption {
     The drop-down buttons ($a$\textasciitilde$d$) allow the user to select the topic, year ranges, and sorting options. ($e$) represents color codes, and ($f$) shows the list of topics.
     } 
    \label{fig_option1}
    \vspace{-0.7cm}
\end{figure}

\subsection{PDViz: Visual Interface}
PDViz consists of three main views: policy diffusion analysis, map, and policy adoption (Fig.~\ref{fig:teaser}-A, B, and C). There are multiple drop-down menus that allow the user to configure PDViz. We developed PDViz using JavaScript including React~\cite{REACTJS} and D3js~\cite{d3js} for the front-end and Flask~\cite{Flask10} for the back-end. We created an R script in the back-end to infer the policy diffusion patterns using the NetworkInference R package. 
Rpy2~\cite{rpy2} allows R scripts to run embedded in our Python3 back-end environment. 
The data flow pipeline for PDViz is shown in Fig.~\ref{fig_pipeline}. The interface is publicly available (\url{https://vizus.cs.usu.edu/app/pdviz}) and see the supplemental video\footnote{\url{https://www.youtube.com/watch?v=jTpOVK1cSAU}}.


\subsubsection{Configuration Menus}
To provide a better understanding of how to use the interface with different measurements and settings, we first explain the configuration menus (Fig.~\ref{fig_option1} and Fig.~\ref{fig_usage_scenario1}(e\textasciitilde h)). Using the menus, the user can specify `Policy Topic,' `Time Period,' and `Analysis Measurement' to analyze diffusion of policy. 
The default values for the policy topic, time period, and analysis measurement are \textit{ALL, 1950-2017}, and \textit{Network Centrality} respectively. We choose the time period from 1950 to 2017 as the default because that period is the post-World War II period when capacity and legislative activity in the states increased in response to citizen demand. The user can specify one of eighteen policy topics (Fig.~\ref{fig_option1}-f) using Fig.~\ref{fig_option1}-a. The user can set the time periods between 1691 to 2017 using the drop-down menus and the slider in Fig.~\ref{fig_option1}-b.

PDViz also provides sorting options for the alignment of states and policies. The default alignment order for the state in the views is ascending alphabetical order, but the user can change it to descending order of the selected \textit{Measurement} using Fig.~\ref{fig_option1}-c. The default order for the policies is also ascending alphabetically, but the user can change it using Fig.~\ref{fig_option1}-d. 
It provides two more options. The \textit{\# of Total Adoption} option sorts the policies in descending order of the total number of adoption. Another option, \textit{Policy Activity} allows the user to customize the list of topics or policy orders by selecting a state in the policy diffusion analysis view. It sorts the topics in descending order of the total number of policy adoption made by the chosen state (Fig.~\ref{fig:teaser}-d). The total adoption indicates the sum of policy creations and existing policy adoption. 

PDViz uses four-color codes: red, blue, purple, and green (Fig.~\ref{fig:teaser}). The blue, red, and purple colors describe the incoming, outgoing, and bi-directional relationships in the inferred policy diffusion patterns.
We also use quartile scales with blue, red, and green colors to represent the number of policy creations, number of existing policy adoptions, and degree of measurement respectively. The values for each quartile are min-max normalized, and each color intensity charges a size of 25\% (Fig.~\ref{fig_option1}-e and ~\ref{fig:teaser}-B).

The user can choose one diffusion analysis method and measurement among the options in Table~\ref{table_analysis_measurement_in_PDViz} using Fig.~\ref{fig_usage_scenario1}-e and -f. When the analysis method is either \textit{Network Centrality} or \textit{Static State Innovativeness}, the range of measurement is limited to the period set in Year on the configuration menu (Years Range in Fig.~\ref{fig_usage_scenario1}-g). When the analysis method is set to \textit{Contextual Factor}, the user can choose one from \textit{All Range, Years Range,} and \textit{1 Year} to change the measurement range. In the case of \textit{All Range}, the measurement range is determined over a total year range between 1680 and 2017. 
When the user selects \textit{1 Year}, an option for changing the basis year becomes visible next to the measurement option (Fig.~\ref{fig_usage_scenario1}-f). When the user changes the basis year, the measurement range is determined by measurements in the year.

PDViz provides a simple search for the user to retrieve where a policy is included in a topic (Fig.~\ref{fig:teaser}-g). When the user types a keyword, it shows sets of topics and policies containing the keyword.

\subsubsection{Policy Diffusion Analysis View}
The policy diffusion analysis view allows the user to explore patterns by state and topic (T1 and T3.) 
It consists of a policy diffusion chart and a policy matrix chart. 
The policy diffusion chart (Fig.~\ref{fig:teaser}-A1) presents the underlying policy diffusion patterns computed using the NetInf algorithm. The policy matrix chart (Fig.~\ref{fig:teaser}-A2) shows the overview of how many new policies were created and existing policies were adopted by state and topic. 
When a topic is specified in the menu, it presents which state created the policies first and which states adopted the policies accordingly (Fig.~\ref{fig:teaser}-A3). 

\textbf{Policy Diffusion Chart:} This chart is designed for the user to explore and identify the policy leaders and what the diffusion patterns look like. It shows two sets of underlying diffusion patterns above and below the state abbreviations with the color-coded glyphs representing the degree of measurements. 
The ties above and below the abbreviations are called the upper and lower patterns respectively in this work. The upper patterns show the patterns for overall policies across the states when a topic is not specified in the menu. Conversely, when a topic is specified, the upper patterns represent the patterns for the topic. The lower patterns represent diffusion patterns at a different level. They are described in the following sections with the policy matrix chart. When the user hovers over a state name or glyph, the upper patterns from the state to others with one of three relations (incoming, outgoing, and both) are highlighted in colors of blue, red, and purple, depending on the relation. 
\\
\textbf{Design Choice.} 
An alternative design is a common 2D network diagram. However, it has a limitation in terms of measurement and pattern comparisons. 
Compared to the network, the arc diagram layout is intuitive to compare the associated values of the US states when they are sorted by the values. It is also beneficial to compare two pattern results by placing them up and down, supporting T3.

\textbf{Policy Matrix Chart:} This shows the heatmap style overview representing how many new and existing policies have been created and adopted by states. The policy matrix and diffusion charts are vertically well-aligned and share the same x-axis (Fig.~\ref{fig:teaser}-d). 
When the topic option is \textit{ALL}, the y-axis displays eighteen topics like Fig.~\ref{fig:teaser}-A2.  
Each cell across the state and topic has blue and red rectangles, and their color intensities represent the relative number of policy creation and existing policy adoption by each row respectively (Fig.~\ref{fig_option1}-e). When the user hovers over a cell, upper and lower diffusion patterns in the policy diffusion chart are highlighted. For example, when the user hovers the mouse over the cell (state: CA, topic: \textit{Law and Crime}), the lower diffusion patterns starting at CA related to the topic are colored in one of blue, red, and purple depending on their relations to CA. At the same time, the upper patterns starting at CA are also highlighted as Fig.~\ref{fig:teaser}-A1. They allow the user to compare the patterns in the topic context (T3).

When the user specifies a policy topic in PDViz, the policy diffusion and policy matrix charts are updated. 
Then, the upper patterns in the policy diffusion chart represent the diffusion patterns for the selected topic, and the lower patterns show the primitive incoming and outgoing connections for a policy between states. Namely, the lower patterns in Fig.~\ref{fig:teaser}-A1 and upper patterns in Fig.~\ref{fig_usage_scenario2}-A for state CA and topic \textit{Law and Crime} are identical. The policy matrix view shows the list of policies in the topic on the y-axis like Fig.~\ref{fig:teaser}-A3. 
When the user hovers the mouse over a policy label, a tooltip appears like Fig.~\ref{fig_usage_scenario2}-B. It shows a list of context factors along with the associated Cox's hazard ratios. By examining these context factors and ratios, the user can find which factor has had the most influence on policy diffusion.
Each row in the policy matrix view has blue and red circles (Fig.~\ref{fig_usage_scenario2}-A). The blue circles are for the candidate states considered as the policy initiator, where they adopted the policy in its first adoption year.
The red circles represent the states that adopted it accordingly. Hovering the mouse over on circles highlights diffusion patterns in the same manner. 

The policy analysis view is coordinated with the other views. When the user hovers the mouse over the states, y-axis labels, and matrix cells, the map and policy adoption views are updated. 
\\
\textbf{Design Choice.}  The design alternative is a small multiples of line chart or bar chart. However, it has two shortcomings compared to the current design. First, it can fully support the user in horizontal comparisons, but observing values vertically is less intuitive. 
Second, it needs more vertical space to represent values in a line or bar that are distinct enough to be noticed for comparison. 

\subsubsection{Map View}
The map view is designed to present measurement information with geographical information. It is closely related to T2, exhibiting the geographic proximity of policy diffusion, such as the neighbor effect. 
It provides two map styles: the general US state map and hexbin map (Fig.\ref{fig_usage_scenario1}-A \& Fig.\ref{fig:teaser}-B). 
Both are the choropleth maps colored in the green quartile (Fig.~\ref{fig_usage_scenario1}-h) by the measurement degrees.

When the user hovers the mouse over a state, the boundary of other states with a connection in the diffusion patterns to the state is highlighted. They are colored in one of three colors (blue, red, or purple) representing the incoming, outgoing, and bi-direction patterns. At the same time, the upper diffusion patterns connected to the state in the policy diffusion view are highlighted, and the policy adoption view presents the filtered results by the state. The map view is pannable and zoomable to allow the user to navigate. 
\\
\textbf{Design Choice.} The hexbin map could be represented in different styles of cartograms~\cite{cano2015mosaic}. 
However, they are ruled out in the design because they do not satisfy the following criteria: 1. presenting measurements in the same size of shapes to avoid misinterpreting the degree of the measurements due to the sizes; 2. keeping the topological accuracy while maintaining adjacencies between the US states. From the list of state-of-art cartograms~\cite{nusrat2016state}, the hexbin map is the best option satisfying both criteria.

\subsubsection{Policy Adoption View}
The policy adoption view has four tabs: By Year, By State, By Topic, and By Context (Fig.~\ref{fig:teaser}-C). The first three tabs provide bar charts showing the number of policies that are newly created or adopted from the other states by year, state, and policy topic. The By Context tab provides a line chart representing social factors and includes small boxes with state abbreviations written on them.

\begin{figure*}[t]
 \includegraphics[width=.95\linewidth]{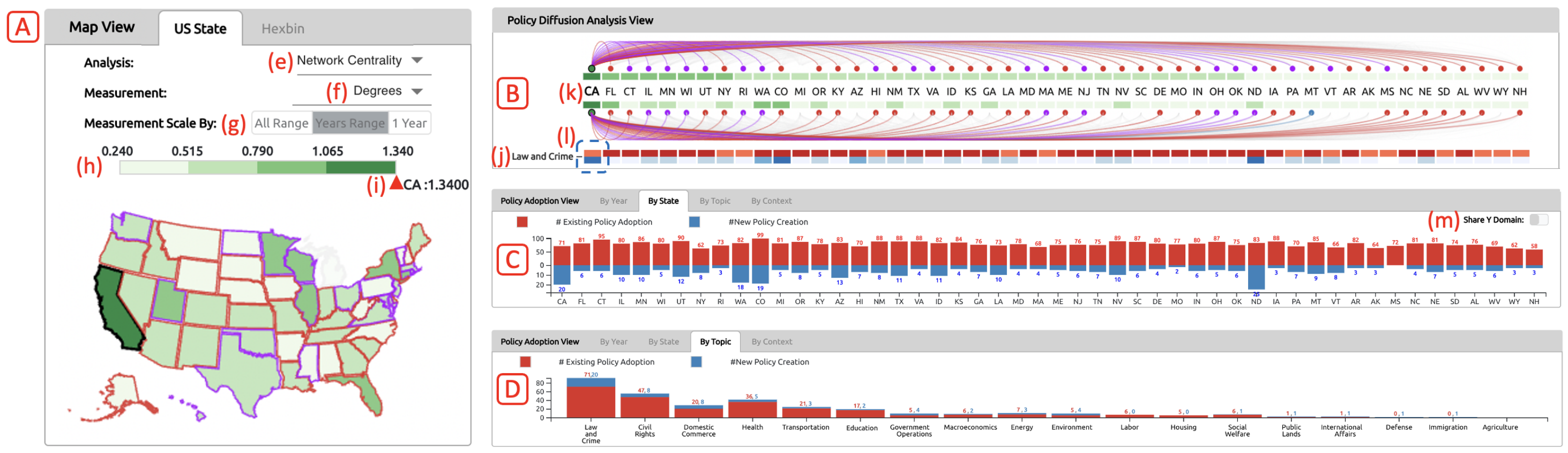}
 \centering
  \caption{The map view (A) provides two styles of maps: the general US state map and hexbin map. The user can change the analysis method and measurement. It also presents the green quartile for the measurements. When the user hovers the mouse over topic labels (j) or state abbreviations (k), the By State (C) and By Topic (D) tab in the policy adoption view show the filtered adoption records by the topic or state.}
\label{fig_usage_scenario1}
\vspace{-0.4cm}
\end{figure*}

\textbf{By Year and State Tabs: }Analyzing the policy adoptions by years and states supports T2. 
The mirror bar charts in the By Year (Fig.~\ref{fig_usage_scenario2}-C) and By State (Fig.~\ref{fig_usage_scenario1}-C) tabs show the number of policies by year and state respectively. 
The blue and red bars indicate the number of policy creation and existing policy adoption. As the number of policy creation is significantly smaller than the number of existing policy adoptions, they provide an option to choose whether to share the domain ranges for the y-axis as shown in Fig.~\ref{fig_usage_scenario1}-m and Fig.~\ref{fig_usage_scenario2}-h. When the user makes mouse-over interactions in the policy diffusion analysis and map views, the tabs show the number of policies filtered by states, topics, or policies. 
\\
\textbf{Design Choice.} 
We choose the bar chart layout to align the categorical variables (i.e., the US states) on the horizontal axis in the same order as the policy diffusion analysis view. When the measurements are represented in a stacked or grouped bar chart, the smaller values could be difficult to identify for the user due to the fact that the number of the policy creation is significantly smaller than the existing policy adoption. As the result, we adopt the mirror bar chart using two different scales up and down for the tabs.

\textbf{By Topic Tab:} 
It (Fig.~\ref{fig_usage_scenario1}-D) is a bar chart showing the number of unique policies by topic, supporting T3. 
When the user hovers the mouse over a state or a cell across a state in the other views, the tab shows the filtered number of policies by the state. The blue and red bars represent the number of policies that the state first introduced and the number of policies which the state adopted from others. 
\\
\textbf{Design Choice.} The design considerations are the same as for the Year and State tabs. However, we use a stacked bar chart because the number of policies by topic is not additionally divided into the two adoption types before specifying a state.



\textbf{By Context Tab:} This tab shows a line-box chart to allow the user to observe policy adoptions along with social factors (T4). Each box represents a state and is located in the year that the state adopted the policy. The number of lines is equal to the number of boxes. There are three drop-down menus as shown in Fig.~\ref{fig_usage_scenario2}-i, from which the user can choose a policy, a state, and a contextual factor to analyze. 
There are three line colors: The green line represents the selected state in the menu, and the black line represents the US average. The gray lines indicate the others. The boxes located in the policy's first adoption year are colored blue, and the other boxes are colored red. When there were multiple adoptions in a year, boxes are stacked from the bottom of the x-axis in ascending order of factor measurements in the year. When the user hovers the mouse over a state box, the line associated with the state is highlighted and shows its details as a tooltip (Fig.~\ref{fig_usage_scenario2}-k). 


In addition, this tab is highly coordinated with the policy matrix chart. When the user clicks on a policy label or a contextual factor in the policy matrix chart, the values in the menu are updated. 
It works as a shortcut to set up the drop-down menus to analyze an interesting policy by a factor.
\\
\textbf{Design Choice.} Visualizations for multiple time series data have many alternative designs, such as a line chart, stacked area chart, and heatmap. However, the stacked area chart and heatmap are excluded from our design choice because they could be difficult to read and compare the values when drawn in a limited space. In addition, they are not visualized with the color boxes representing when the states adopted a policy in the tab.

\vspace{-0.3cm}
\section{Usage Scenarios}
\label{sec_usage_scenario}

\subsection{Understanding Policy Diffusion and State Roles}
Bob, a policy analyst, is interested in answering questions such as \textbf{(T1)}: ``Q1. What role do various US states play in policy diffusion processes?,'' ``Q2. What are the underlying patterns for a state in policy diffusion?,'' and ``Q3. What are the statistics of adopted policies by states?''; \textbf{(T2)}: ``Q4. How does the states' role in policy diffusion change by year?''; \textbf{(T3)}: ``Q5. How are the underlying diffusion patterns different by policy topic?'' PDViz provides Bob with multiple analysis methods and interactions to find the answers he has from different perspectives 
(see the supplemental video\footnote{\url{https://www.youtube.com/watch?v=Ezi\_GDcWXmA}}). 

\textbf{Q1 \& Q4:} Bob identifies the measurements of `Network Centrality' and `Static State Innovativeness'. He gives \textit{Measurement} to the `State Sort Option' (Fig.~\ref{fig_option1}-c) to rearrange the state order in the policy diffusion analysis view (Fig.~\ref{fig_usage_scenario1}-B and -C). 
It allows him to read the US states' roles and impacts easily by changing the analysis measurements:
\textit{\textbf{Closeness}} shows how impactful a state's role is as a policy leader in disseminating policies across the states. Colorado (CO), Arizona (AZ), and Illinois (IL) are some of the quickest influencers in the underlying policy diffusion. 
\textit{\textbf{PageRank}} informs that the state's importance or popularity in the policy diffusion. Georgia (GA) is the most important state, followed by AZ, and Louisiana (LA). \textit{\textbf{Out-Degree}} tells how much direct impact a state has had on other states. In other words, the out-degree describes which states created the most influential policies. CA has affected the others most, followed by Florida (FL) and Connecticut (CT). Conversely, \textit{\textbf{In-Degree}} describes how much direct impact a state has received from others. It refers to the extent to which a state has accepted policies developed by other states. CA also has been influenced by the others most, and CT, and FL are followed. \textit{\textbf{Degree}} is a combination of the out-degree and in-degree centralities. CA has been actively affected or being influenced by the other states the most, and FL, and CT are followed. \textit{\textbf{Static State Innovativeness}} tells Bob a state's willingness to adopt policies sooner or later than others. He finds Oregon (OR), New Mexico (NM), and CO have been the fastest early adopter of policies. By changing the year range, he could find different roles of states in the policy diffusion to answer \textbf{Q4}. 


\textbf{Q2:} Bob observes policy diffusion patterns by hovering the mouse over a state in the policy diffusion view.
He chooses CA to analyze as it plays a key role in three centralities. 
CA has the red ties to 29 states including FL, New York (NY), and Montana (MT), and has the purple ties to 19 states including CT, IL, and Washington (WA) (Fig.~\ref{fig_usage_scenario1}-k). 
The red ties tell Bob that when CA adopts a policy, 29 states have a high probability to adopt the policy. The purple ties indicate that CA has a pattern that when one of 19 states adopt a policy, CA probably adopts the policy and vice versa. He found that CA has no blue pattern and no tie to Michigan (MI).


\textbf{Q3:} Bob opens the By Topic tab in the policy adoption view. He discovers that the US states have a high interest in \textit{Law and Crime} and \textit{Civil Rights} because those topics have more policies than others from 1950 to 2017. Then, he hovers the mouse over on CA to identify the statistics of policy adoptions that CA has made. He finds that CA also has the similar statistics, adopting the most \textit{Law and Crime} policies, followed by \textit{Civil Rights} policies (Fig.~\ref{fig_usage_scenario1}-D). He identifies further that, of the 91 \textit{Law and Crime} policies, CA is the potential creator of the 20 policies. 


\textbf{Q5:} The policy diffusion view allows Bob to compares the differences between overall and topic-specific diffusion patterns. Bob hovers the mouse over the cell across CA and \textit{Law and Crime} (Fig.~\ref{fig_usage_scenario1}-l), and compare the upper and lower patterns in the policy diffusion chart. CA has the common underlying outgoing patterns to FL and NY, and the outgoing pattern to MT is changed to the incoming pattern in \textit{Law and Crime}. Bob becomes interested in what makes the incoming pattern between MT and CA in \textit{Law and Crime}. In the configuration menu, he specifies the topic \textit{Law and Crime}. Then he sets the \textit{Policy Activity} to the `Policy Sorting Option' and customizes `Policy Order' in the policy matrix chart by clicking CA. He identifies CA adopted 6 \textit{Law and Crime} policies from MT such as \textit{Protection of Victims of Sex Crimes during Criminal Proceedings} and \textit{Retail Theft Law}. He concludes that these 6 policy adoptions from MT to CA in \textit{Law and Crime} disclose the hidden pattern in the overall pattern.

\begin{figure*}
  \centering
    \includegraphics[width=.95\linewidth]{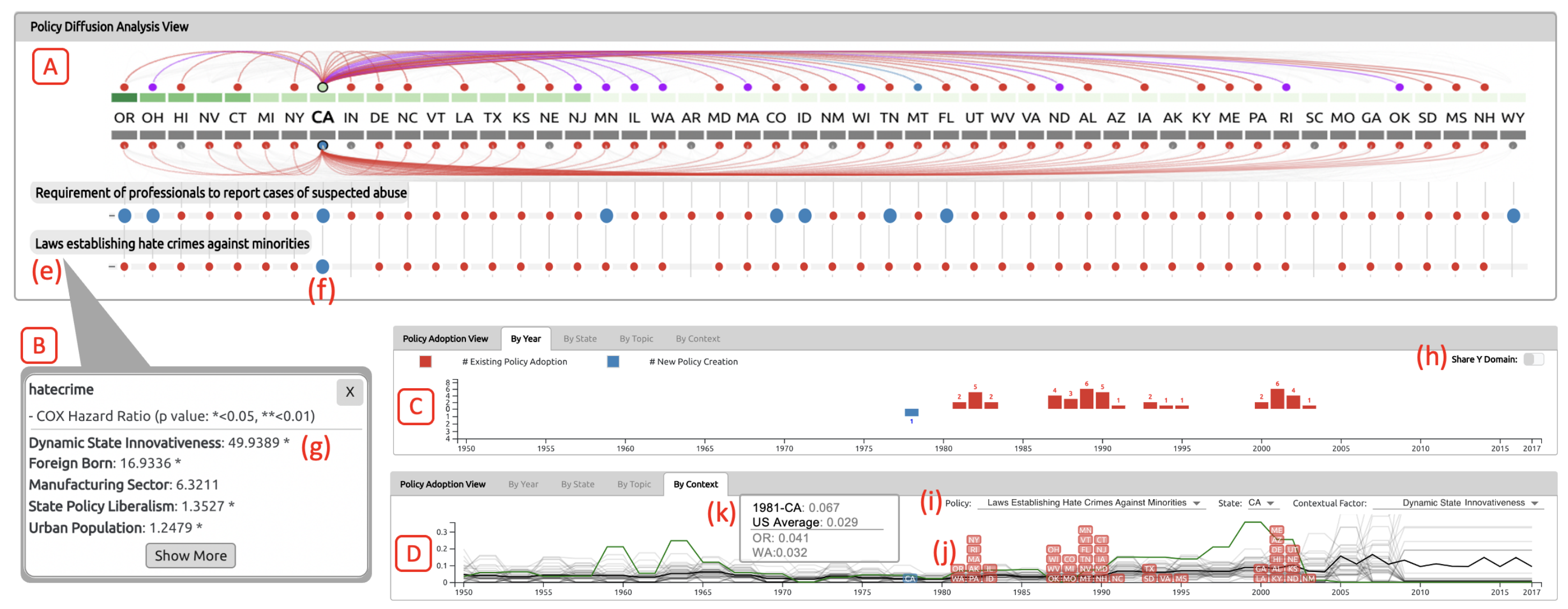}
     \caption {
    When the user makes a mouse-over interaction on a policy label (e), it shows a tooltip (B). It shows the list of the contextual factors along with their Cox's hazard ratios. The By Year (C) and By Context tabs (D) show policies' adoption trend.  
     } 
    \label{fig_usage_scenario2}
    \vspace{-0.6cm}
\end{figure*}

\vspace{-0.3cm}
\subsection{Policy Adoption and Social Factor}
Cora, a journalist, is interested in minority hate crimes. Using PDViz, she can answer questions like \textbf{(T1):} ``Q1. Which state enacted a policy and which states have adopted the policy accordingly?'' and ``Q2. When did the states adopt the policy?''; \textbf{(T4):} ``Q3. Which factors affect the states to adopt the policy?''

\textbf{Q1:} She first searches her keyword `minorities' and finds her target policy \textit{Laws establishing Hate Crimes against Minorities} is included in the topic \textit{Law and Crime}. Cora sets the topic in the configuration menu and finds the target policy as shown in Fig.~\ref{fig_usage_scenario2}-e. She easily identifies the policy was first adopted by CA as there is only one blue circle in its row (Fig.~\ref{fig_usage_scenario2}-f). She also discovers that most states have adopted the policy except the 4 states (Arkansas (AR), Indiana (IN), South Carolina (SC), and Wyoming (WY)).

\textbf{Q2:}
The By Context tab (Fig.~\ref{fig_usage_scenario2}-D) helps Cora track when the US states adopted the minority policy and which states adopted the policy earlier than others. She clicks the policy label in the policy matrix chart to set the menus in the tab (Fig.~\ref{fig_usage_scenario2}-i). She finds that CA adopted the minority hate crimes policy in 1978 first. Next, she finds 9 states, its early adopters. WA and OR adopted the policy in 1981, and NY, Rhode Island (RI), Massachusetts (MA), Alaska (AK), and Pennsylvania (PA) adopted the policy in 1982. Idaho (ID) and IL adopted it in 1983. 

\textbf{Q3:} The tooltip in the policy diffusion analysis view (Fig.~\ref{fig_usage_scenario2}-B) and the By Context tab allows Cora to explore what contextual factors affect the states to adopt the policy faster than other states.  The tooltip displays a list of context factors along with the associated Cox' hazard ratios. She identifies that \textit{Dynamic State Innovativeness} has the biggest impact (49.9389), followed by \textit{Foreign Born} (16.9336). The * marker is placed next to the values for each indicating that the factor has a strong relationship to policy adoption ($p$\textless 0.05). 
She investigates the \textit{Dynamic State Innovativeness} scores in their adoption years first. CA had .041 which was a higher value than the US average (.022) in 1978. OR (.041) and WA (.032) in 1981, NY (.071), RI (.064), MA (.061), AK (.050), and PA (.047) in 1982, and IL (.105) in 1983 also had the higher values than the US average of their adopted years (.029 in 1981 and 1982, and .042 in 1983) respectively. However, ID (.029) in 1983 had a smaller measurement than the average. Next, she changes the contextual factor to \textit{Foreign Born} to examine. She finds CA had a population that was 10.5\% foreign-born, and it was higher than the US average (6.0\%) in 1987. However, she finds that WA (6.8\%) and OR (4.7\%) in 1981 and ID (3.0\%) in 1983 had smaller measurements than the US average each (7.2\% in both 1981 and 1983). 

She discovers that WA and OR adopted the minority policy earliest. She also finds that ID adopted the policy despite the fact that both factors were much lower than the US average. She becomes curious as to what other factors influenced them to adopt the policy so early. In the map view, she notices that they are geographically close to CA. The second early adopter group (NY, RI, MA, and PA) is geographically close to each other as well. She concludes that \textit{Dynamic State Innovativeness} and \textit{Foreign Born}, as well as geographical proximity, influenced the adoption of the minority policy.

\vspace{-0.3cm}
\section{Expert Feedback}
To evaluate the exploratory capabilities and usability of PDViz, we have conducted a user study (IRB-approval \#12187) with two political scientists (E1 and E2).  
They are the experts in American politics and political methodology and have long been doing research on NetInf and SPID. 
They hold a PhD degree in political science and have taught courses in the department of political science at universities having 15 and 14 years of experience. E1 has experience with data visualization tools which are developed using Tableau~\cite{tableau} and R shiny~\cite{shiny}. 
His self-reported familiarity with the visualization tools is 4 out of 7. E2 has no experience in using visualization interfaces before. The experts were not involved in the design and implementation of PDViz. 

\vspace{-0.3cm}
\subsection{Interview Setup \& Procedures}
Upon the domain experts agreeing to participate in the user study, we conducted a one-hour introduction session with them. At the session, we introduced PDViz and demonstrated the two usage scenarios, the same as those described in Section~\ref{sec_usage_scenario}. During this session, they learned how to use PDViz. 

We sent a guideline detailing the step-by-step instruction to complete the study after the session ended. It includes the recorded session video and text descriptions introducing PDViz and the two usage scenarios. 
They attended the study individually in a virtual setup. 
They answered a set of pre-questionnaire asking about their demographic information and their academic qualification like experience in political science. Next, they followed the usage scenarios using PDViz themselves and freely use it for 10-20 minutes. They completed the post-questionnaire at last. The post-questionnaire asked them to rate PDViz on a 7-point Likert scale (1, strongly negative to 7, strongly positive) based on their user experience, such as how helpful each view was and how much T1-T4 are supported by PDViz. They also listed features that they would like to see added to PDViz.

\vspace{-0.3cm}
\subsection{Feedback Results}
The experts positively assessed PDViz, acknowledging that it provides results that align with their domain knowledge (see the expert review\footnote{\url{http://vizus.cs.usu.edu/downloads/pdviz/expert\_review.pdf}}). E1 reported that PDViz allows him to understand systematic variation in diffusion patterns across states and how those patterns change over time. E1 discussed that one state might tend to follow another state with a policy diffusion tie, but also pointed out that users should keep in mind that diffusion does not necessarily imply causality. He also strengthened the potential usage of PDViz as a teaching tool, especially in undergraduate courses as the concept of policy diffusion is not intuitive to all students. E2 is interested in policy diffusion in general, and he agreed that the policy topics that PDViz provides are well relevant to his domain interests. He reported that the combination of domain knowledge and PDViz allows him to understand the diffusion network structure.

Overall, the post-questionnaire results show that PDViz is well designed and fairly easy to use. They liked the use of multiple views. The experts reported that PDViz successfully provides an overview of the policy diffusion (T1) giving scores of 7, and 6. Both experts agreed that PDViz supported them to explore the policy diffusion in geo-spatial and temporal context (T2) giving 7. They rated 7, and 5 scores for PDViz to analyze the diffusion in topical context (T3), and rated 7, and 4 to analyze it in a social context (T4).  

The experts offered further comments on the PDViz interface. E1 liked the policy analysis view the most as it summarizes a lot of information efficiently. He also reported that he likes the use of the colors which condense information well. Otherwise, E2 liked the map view the most since it is easy to see aggregate patterns via the maps which are familiar to him. He reported that PDViz is very interesting and helpful to explore the diffusion patterns but pointed out that the illustration of the diffusion networks could be improved as the ties in the policy diffusion chart are hard to see. Both reported that it would be interesting for users to add data and update the views in PDViz, and download the results used in constructing a specific view. E1 appended that PDViz provides comprehensive patterns but it would be more interesting if he can analyze how the policy diffusion changes over time at once. 

\vspace{-0.3cm}
 \section{Discussion}

PDViz can be utilized by policymakers and political scientists for exploratory purposes around broad policy movement across the states, or for specifics around an individual policy.
PDViz adheres to the visualization mantra of providing an overview first (T1), and the user can intuitively explore details at the state level or policy domain based on their interests via simple mouse point-and-click actions (T2). In the user study, the political scientists were asked to use the PDViz interface after watching a simple introduction video. They rated that PDViz is easy to use. Therefore, PDViz is expected to have a short learning curve for potential users even if the users are non-visualization experts. 

PDViz is also interested in informing and providing evidence for policy decisions across the state. The policy diffusion patterns indicate that a common policy spread. However, some states are political leaders who are innovators of new policy, and some are laggards. Policymakers look at other states to learn, emulate, or compete (T3). PDViz could partially support them to investigate what similar economic, social and political circumstances have influenced their policy adoptions (T4) as Section~\ref{sec_usage_scenario}. To fully support their tasks, we report the limitations and future work based on the domain experts' feedback as follows.

\textbf{Data and Scalability:} 
PDViz is a data-driven interface and uses the US demographic and policy data, especially SPID and CSPP. PDViz has a potential to be extended by replacing the policy history data and adding further layers of social factors. We may consider applying different policy and policy diffusion data observed from other countries such as the United Kingdom and Italy with updating the map view. It could be achieved by adding a function to upload the data into the interface, as the domain experts pointed out. 


\textbf{Policy Adoption Pattern Comparisons:} 
As E1 pointed out, the user might be interested in comparing the diffusion pattern changes over time. Comparing the patterns between more than two different topics or policies could also be of interest. To our best knowledge, none of the previous visualization systems was designed for political scientists (e.g., PolicyFlow) to support those tasks. Allowing the user to perform the tasks is the focus of our future research. 


\textbf{Machine Learning Model and Input Parameters:} We utilize machine learning algorithms such as the Cox model and the NetInf algorithm. The output of these models is influenced by the input parameters. 
We used the default parameters in this work. Although no comments from the experts were reported regarding the parameters, the user could find different results and insights with the different input parameters. 
In future work, we plan to add methods that allow the user to change the input parameters within PDViz and to evaluate what interpretations are available.

\vspace{-0.3cm}
\subsection{Lesson Learned} 

\textbf{L1. Visualization System as a Teaching Tool.} PDViz is designed for domain experts in political science to support their policy data exploration and diffusion analysis. Interestingly, the domain experts remarked that the visual tie metaphor in PDViz could help students understand the fundamentals of policy diffusion and the role of states in policy diffusion. It refers to a potential use of visualization techniques for educational purposes. Understanding the effectiveness and impact of PDViz on social science students is a direction of future work and will help us better understand the potential of visualization approaches as a teaching tool.

\textbf{L2. Color Use.} The red and blue colors are widely used to refer to political parties like Republicans and Democrats. PDViz uses these colors and additional purple to represent different policy diffusion patterns. For our color choice, the domain experts noted that they condense information well and are useful for understanding policy diffusion patterns. It shows that the reasonable use of colors could help the domain experts do their analysis works.

\textbf{L3: Policy Diffusion does not necessarily imply causality.} Policy diffusion, as a result of NetInf, presents patterns of how the policy is diffused across the states and used as predictors. 
However, as the domain experts noted, the diffusion ties refer to the tendency but they do not take account into for causality by themselves. It also does not explain mechanisms of policy adoption, such as whether the US states have adopted existing policies or created new policies that are similar to existing ones in competition with other states to cope with similar social and political situations. Providing insights into causalities and mechanisms is our future research.

\textbf{L4: Domain Specific Lesson.} This lesson is for researchers and practitioners who consider visualizing policy diffusion networks between countries or states. We found that more than one state possibly created a policy (or similar policies). This is because the policy's historical records (e.g., SPID) do not have related information (the issue could be clarified in future updates.) The Cox model estimates the impacts of the covariates represented in the hazard ratios. However, unless presented with $p$-values, it is impossible to tell whether the impacts are significant or not. Therefore, the hazard ratios should be presented together with their $p$-values.



\vspace{-0.3cm}
\section{Conclusion}

We introduced PDViz, a visual analytics approach that enables political scientists to easily identify and analyze state-level policy data. It has multiple views to help users explore policy diffusion events in terms of geospatial, policy topic area, and demographic context. We evaluated the interface by presenting two usage scenarios that show how the user can utilize it for analytical tasks and reporting the domain experts' feedback. The domain experts reported positive feedback that PDViz is fairly easy to use and helps them analyze policy diffusion patterns across US states and the context factors. They also reported that PDViz provides the results that are consistent with their domain knowledge. Furthermore, they suggested the features that they would like to see added to PDViz and encouraged us to study a visualization for policy diffusion changes over time as future work.



\printbibliography

\end{document}